\documentclass[twocolumn,prl,amsmath,letterpaper,floatfix,showpacs]{revtex4}
\usepackage{graphicx}
\usepackage{amsmath}
\usepackage{amssymb}
\usepackage{color}
\usepackage{textcomp}

\newcommand{\Ntwo}{{N$_{2}$ }}

\newcommand{\Tcol}{\tau_\text{c}}
\newcommand{\Trot}{T_\text{J}}

\begin{document}
\title{Effects of ultrafast molecular rotation on collisional decoherence}

\author{Alexander A. Milner, Aleksey Korobenko, John W. Hepburn and Valery Milner}

\affiliation{Department of Physics \& Astronomy, University of British Columbia, 6224 Agricultural Road, Vancouver, BC, Canada V6T 1Z1.}

\begin{abstract}
Using an optical centrifuge to control molecular rotation in an extremely broad range of angular momenta, we study coherent rotational dynamics of nitrogen molecules in the presence of collisions. We cover the range of rotational quantum numbers between $J=8$ and $J=66$ at room temperature and study a cross-over between the adiabatic and non-adiabatic regimes of rotational relaxation, which cannot be easily accessed by thermal means. We demonstrate that the rate of rotational decoherence changes by more than an order of magnitude in this range of $J$ values, and show that its dependence on $J$ can be described by a simplified scaling law.
\end{abstract}

\maketitle

Rotational decoherence in dense gaseous media is an area of active research because of its importance in the fundamental understanding of the dissipative properties of gases, as well as in the practical aspects of thermochemistry and combustion research\cite{Martinsson93, Thumann97, Seeger09, Miller11a}. Laser control of molecular rotation has been successfully applied to numerous physical and chemical processes, in which long lived rotational coherence is essential (for a comprehensive review, see Ref.\citenum{Seideman05}).

One of the most interesting aspects of the collision-induced rotational decoherence is the question about its dependence on the speed of molecular rotation and temperature. From the very first experimental works on the topic \cite{Polanyi72, Brunner79}, it was suggested that the rate of rotational relaxation should drop with increasing rotational quantum number $J$, i.e. that the faster molecular rotors are more robust with respect to collisions. This expectation stems from the intuitive ``exponential-gap law'' (EGL) according to which the decay rate decreases as $\exp[-\Delta E_{J}/k_B T]$ with the increasing distance between the rotational levels $\Delta E_{J}$ (here $k_B$ is the Boltzmann constant and $T$ is the temperature of the gas). The refined version of EGL, known as the ``energy corrected sudden'' (ECS) approximation and introduced by DePristo \textit{et al.}\cite{Depristo79}, is a popular model which successfully explained a large number of experimental observations\cite{Brunner81, Koszykowski87, Martinsson93, Thumann97, Miller11, Kliewer12}.

The ECS theory describes the collisional decay rate in terms of an ``adiabaticity parameter'' $a \equiv \omega _J \Tcol = \omega _J l_c / v_c$, where $\omega _{J}$ is the frequency of molecular rotation, $\Tcol$ is the collision time, $l_c$ is a characteristic interaction length (usually determined empirically) and $v_c$ is the mean relative velocity between the collision partners. Since $a = 2\pi \Tcol/\Trot$ (with $\Trot$ being the rotation period), it may also be viewed as the angle, by which a molecule rotates during the collision process. When $a \ll \pi$, the collision is sudden and the energy transfer does not depend on $J$. In the case of the finite duration of collisions, i.e. $a \gtrsim \pi$, the ECS model calls for scaling the decay rate with a $J$- and temperature-dependent correction factor
\begin{equation}\label{Eq_AdiabatictyFactor}
    \Omega_{l_c,v_c}(J) \equiv \left[ (1+ a^2/6) \right] ^{-2},
\end{equation}
with `1/6' being specific to $R^{-6}$ interaction potentials.

In thermal ensembles, both the highest available $\omega _{J}$ and the mean particle velocity $v_{c}$ scale equally with temperature. Hence, thermally accessible values of $a$ do not increase with increasing $T$ and typically stay below the adiabaticity threshold, $a \approx \pi $ (see quantitative analysis later in the text). As a result, extensive studies of collisional line broadening in spontaneous\cite{Berard83}, stimulated\cite{Millot92} and coherent anti-Stokes Raman scattering\cite{Martinsson93, Thumann97, Miller11a, Kliewer12}, though successful in reaching relatively high values of angular momentum (e.g. $J=42$ in nitrogen in Ref.\citenum{Kliewer12}), fall short of testing various scaling laws in the adiabatic regime of collisional decay.

In 2005, Ramakrishna and Seidemann\cite{Ramakrishna05} suggested the use of molecular alignment, induced by ultrashort laser pulses, as an alternative approach to studying the decay of rotational coherence in dissipative media. First reported in Ref.\cite{Morgen93}, this method was further developed in a series of experimental works\cite{Vieillard08, Owschimikow10, Vieillard13} and the results were successfully explained by the ECS approximation \cite{Hartmann12, Hartmann13}. However, exciting high rotational states with ultrashort pulses is rather difficult because of the detrimental effect of the strong-field photo-ionization which prevents one from crossing the adiabaticity threshold of rotational decoherence.

Here, we utilize the method of an ``optical centrifuge''\cite{Karczmarek99, Villeneuve00}, recently extended by our group to enable control of the created rotational wave packets\cite{Korobenko14a}, for studying collisional decoherence as a function of the degree of rotational excitation. By spinning nitrogen molecules in extremely broad range of $J$ numbers (from thermally populated $J\approx8$ to ``super rotation'' with $J>60$), we observe more than an order-of-magnitude change in the decoherence rates. Since in our case the control of the molecular angular momentum is executed separately from changing the gas temperature, we explore the molecular dynamics at the cross-over between the non-adiabatic and adiabatic regimes of collisional relaxation, testing the ECS model beyond the adiabaticity threshold.

\begin{figure}[tb]
\includegraphics[width=.75\columnwidth]{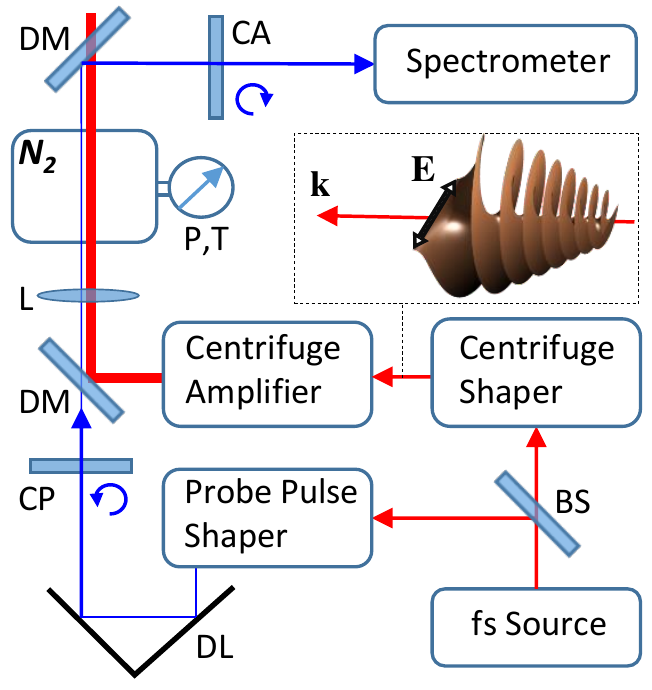}
\caption{Experimental set up. BS: beam splitter, DM: dichroic mirror, CP/CA: circular polarizer and analyzer of opposite handedness, DL: delay line, L: lens. `N$_{2}$' marks the pressure chamber filled with nitrogen gas under pressure $P$ and temperature $T$. An optical centrifuge field is illustrated above the centrifuge shaper with \textbf{k} being the propagation direction and \textbf{E} the vector of linear polarization undergoing an accelerated rotation.}
\label{Fig-Setup}
\end{figure}
The experimental setup is similar to that used in our original demonstration of molecular super rotors\cite{Korobenko14a}. As shown in Fig.\ref{Fig-Setup}, a beam of femtosecond pulses from an ultrafast laser source (spectral full width at half maximum (FWHM) of 30 nm) is split in two parts. One part is sent to the ``centrifuge shaper'' which converts the input laser field into the field of an optical centrifuge according to the original recipe of Karczmarek \textit{et al.} \cite{Karczmarek99, Villeneuve00}. Centrifuge pulses are about 100 ps long, and their linear polarization undergoes an accelerated rotation, reaching the angular frequency of 10 THz by the end of the pulse. The centrifuge shaper is followed by a home built Ti:Sapphire multi-pass amplifier boosting the pulse energy up to 50 mJ. The second (probe) beam passes through the standard $4f$ Fourier pulse shaper\cite{Weiner00} employed for narrowing the spectral width of probe pulses (similarly to Ref.\citenum{Miller11a}) down to 0.06 nm (FWHM, 3.75 cm$^{-1}$) around the central wavelength of 398 nm (achieved by frequency doubling).

\begin{figure}[tb]
\includegraphics[width=1\columnwidth]{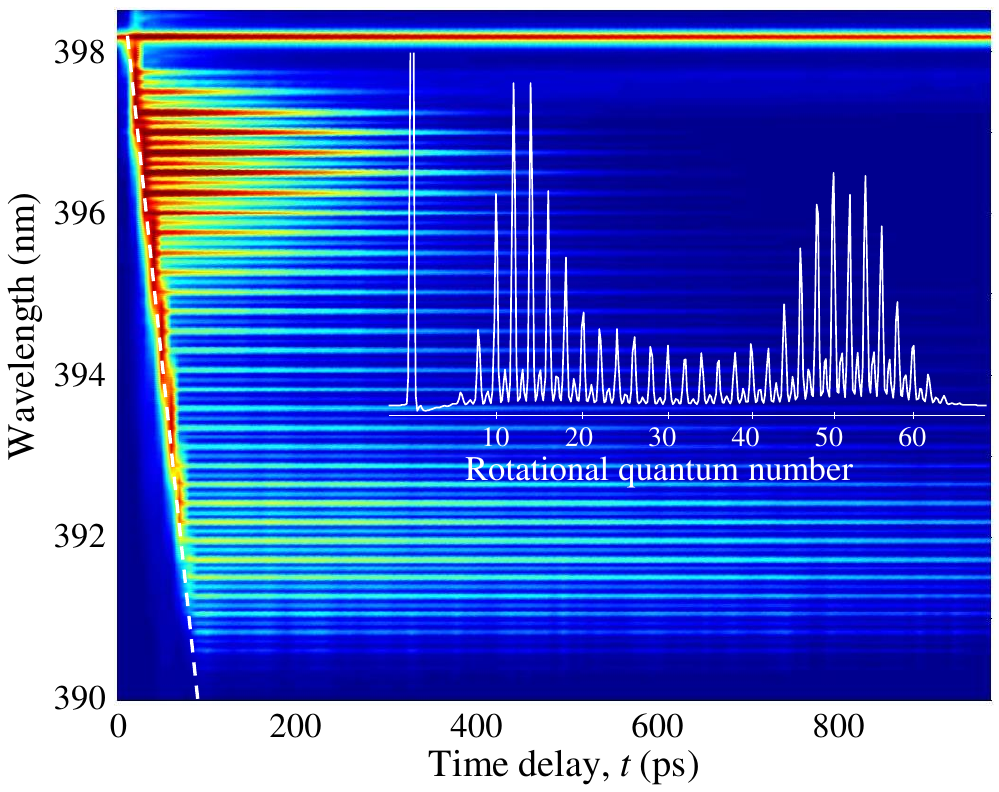}
\caption{Experimentally detected Raman spectrogram showing the rotational Raman spectrum as a function of the time delay between the beginning of the centrifuge pulse and the arrival of the probe pulse. Color coding is used to reflect the signal strength. Tilted white dashed line marks the linearly increasing Raman shift due to the accelerated rotation of molecules inside the 100 ps long centrifuge pulse. A one-dimensional cross section corresponding to the Raman spectrum at $t=270$ ps is shown in white.}
\label{Fig-2DSpectrogram}
\end{figure}
After being combined on a dichroic mirror, centrifuge and probe pulses are focused by a 1 m focal length lens into a chamber filled with nitrogen gas at variable pressure and temperature. To avoid ionization, the intensity of the centrifuge beam is kept below $5\times10^{12}$ W/cm$^{2}$, whereas the probe intensity is more than four orders of magnitude weaker. As demonstrated in our previous work\cite{Korobenko14a}, the centrifuge-induced coherence between the states $|J,m=J\rangle$ and $|J+2,m=J+2\rangle$ (where $m$ is the projection of $\vec{J}$ on the propagation direction of the centrifuge field) results in the Raman frequency shift of the probe field. From the selection rule $\Delta m=2$ and the conservation of angular momentum, it follows that the Raman sideband of a circularly polarized probe is also circularly polarized, but with an opposite handedness. Due to this change of polarization, the strong background of the input probe light can be efficiently suppressed by means of a circular analyzer, orthogonal to the input circular polarizer (CA and CP, respectively, in Fig.\ref{Fig-Setup}).

The Raman spectrum of the probe pulses scattered off the centrifuged molecules is measured with an f/4.8 spectrometer equipped with a 2400 lines/mm grating as a function of the probe delay relative to the centrifuge. An example of the experimentally detected Raman spectrogram is shown in Fig.\ref{Fig-2DSpectrogram}. It reflects the accelerated spinning of molecules inside the centrifuge during the first 100 ps (marked by a tilted dashed white line). While spinning up, the molecules are ``leaking'' from the centrifuge, producing a whole series of Raman sidebands - a set of horizontal lines shifted from the probe central wavelength of 398 nm. Narrow probe bandwidth (smaller than the line separation of 8 cm$^{-1}$ by a factor of 2) enables us to resolve individual rotational states and make an easy assignment of the rotational quantum numbers to the observed spectral lines. This is demonstrated by the Raman spectrum taken at $t=270$ ps and shown in white. The created wave packet consists of a large number of odd and even $J$-states, corresponding to para- and ortho-nitrogen, respectively, whose 1:2 relative population ratio explains the observed alternation of amplitudes.

\begin{figure}[tb]
\includegraphics[width=1\columnwidth]{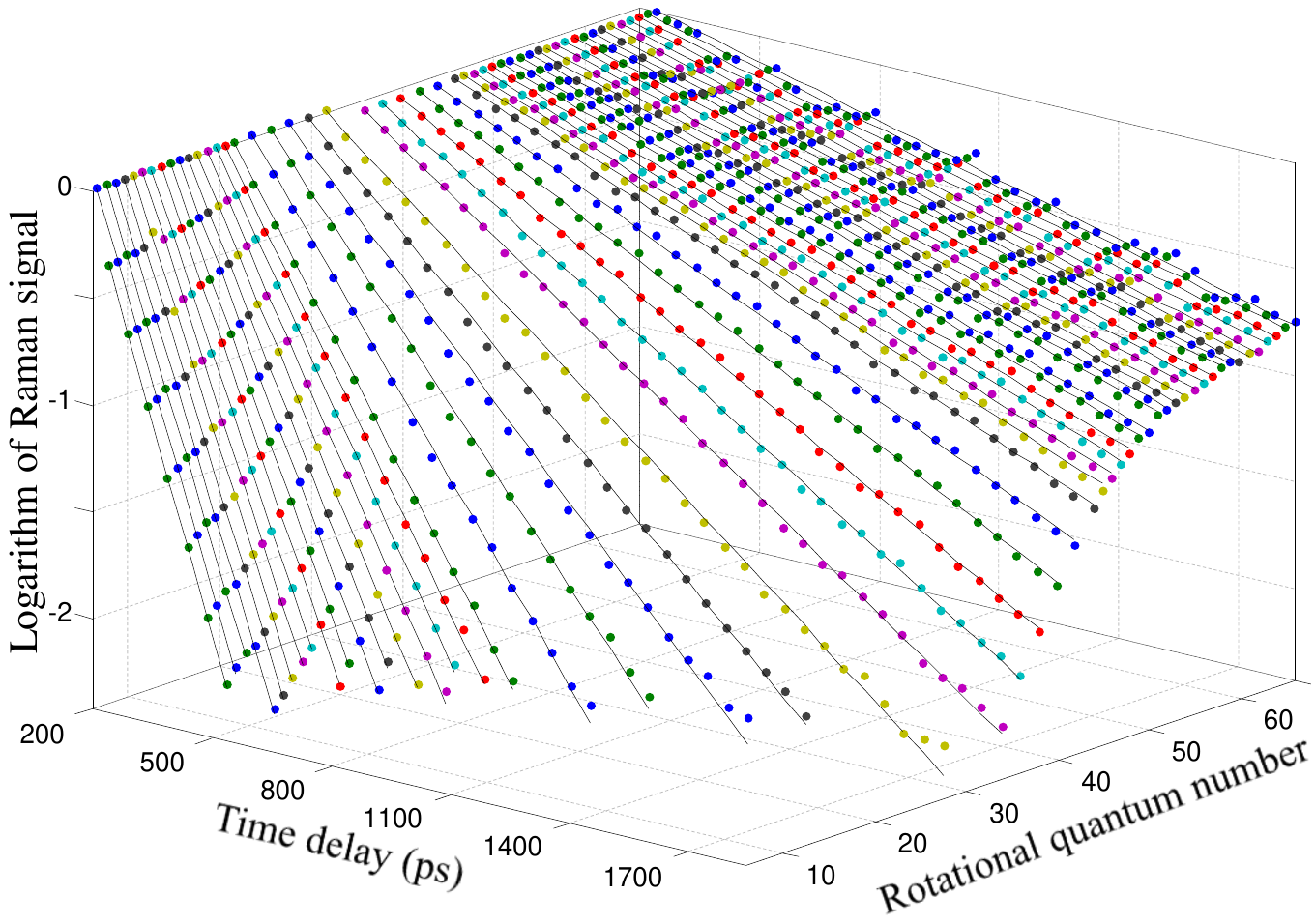}
\caption{Logarithm of the intensity of the experimentally measured Raman lines as a function of time (normalized to 1 at $t=200$ ps, chosen so as to avoid the effects of the detector saturation at earlier times). Dots of the same color represent experimental data for one particular value of the rotational quantum number. Black solid lines show the numerical fit to the corresponding exponential decay. Data collected at $P=0.75$ atm and $T=294$~K.}
\label{Fig-3DDecay}
\end{figure}
The light intensity of each Raman line is proportional to the square of the rotational coherence $\rho _{J,J+2}$, while the frequency shift equals the frequency difference $\omega _{J,J+2}$ between the rotational levels separated by $|\Delta J|=2$. As one can see in Fig.\ref{Fig-2DSpectrogram}, Raman lines corresponding to higher values of angular momentum decay slower than those with lower $J$'s. The decay is happening on the time scale of hundreds of picoseconds, much longer than the duration of our probe pulses ($\approx 4.5$ ps). This offers the possibility to analyze the decay of rotational coherences with both state and time resolution. We plot the time dependence of the intensity of 50 Raman peaks on a logarithmic scale in Fig.\ref{Fig-3DDecay} (colored dots). The observed decay of each Raman line is well described by a single exponential decay, in agreement with a simple theory of decoherence due to random binary collisions.

\begin{figure}[tb]
\includegraphics[width=1\columnwidth]{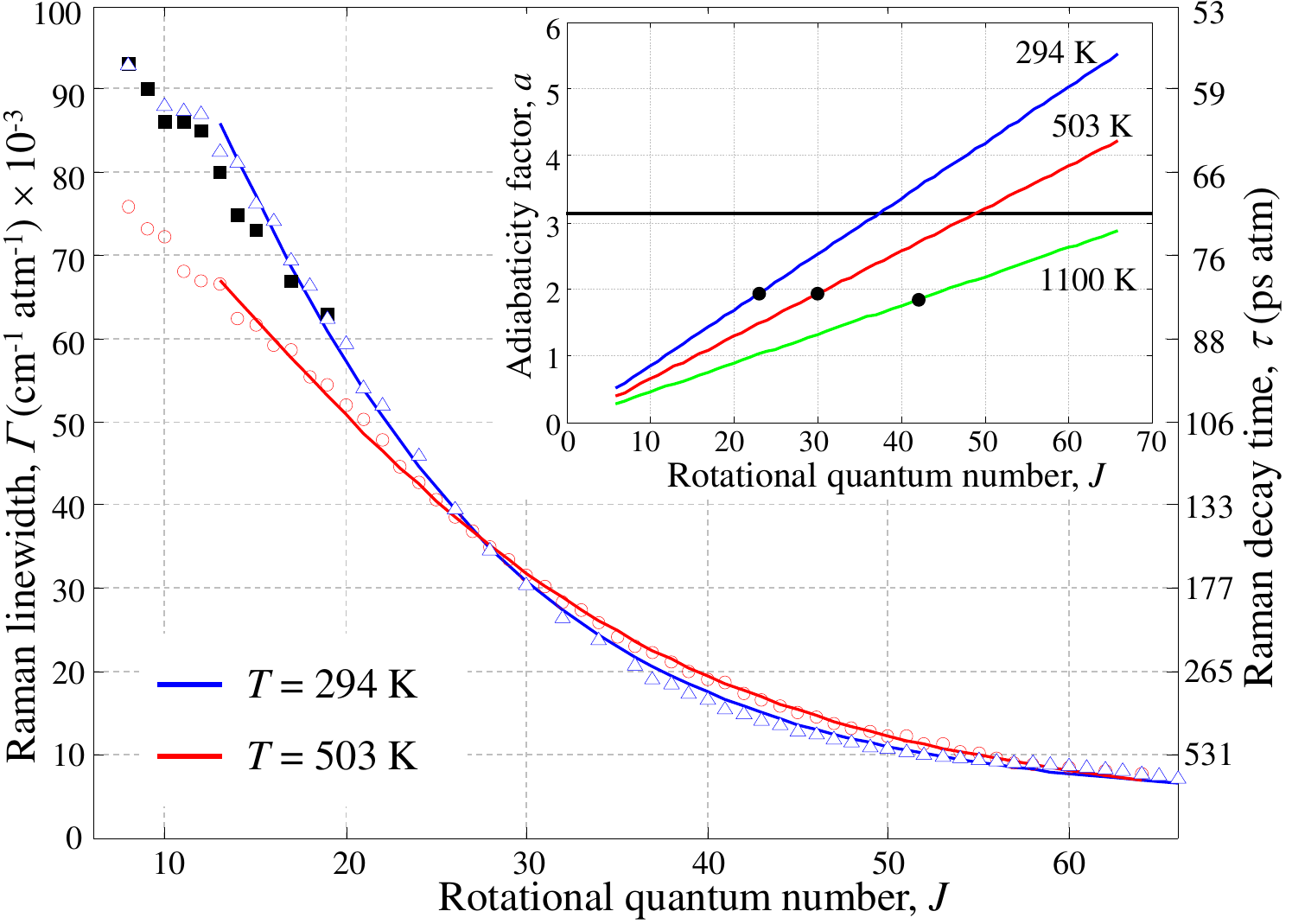}
\caption{Decay rate of rotational coherence of \Ntwo as a function of the rotational quantum number at two different temperature values, $T=294$~K (blue triangles) and $T=503$~K (red circles), expressed as the Raman linewidth (see text). Solid curves correspond to the prediction of the simplified ``energy corrected sudden'' (ECS) approximation at $J>12$. Black squares represent the data from Ref.\citenum{Miller11}. The dependence of the adiabaticity factor $a$ on $J$ is shown in the inset for both temperatures, as well as for $T=1100~K$. Solid horizontal line marks the adiabaticity threshold, $a =\pi$. Black filled circles mark the rotational levels containing 0.2\% of the total population at a given temperature.}
\label{Fig-1DDecay_ECS}
\end{figure}
The dependence of the exponential decay rate on the rotational quantum number at room temperature is shown by blue triangles in Fig.\ref{Fig-1DDecay_ECS}. Expressed in the often used units of Raman linewidth (cm$^{-1}$ atm$^{-1}$), it is calculated as\cite{Miller11, Kliewer12}:
\begin{equation}\label{Eq_RamanLinewidth}
    \Gamma_{J} = \left(  2\pi c \tau_{J} \right) ^{-1},
\end{equation}
where $\tau_{J}$ is the exponential decay time of the Raman signal  corresponding to the transition between states $J-2$ and $J$ (right vertical scale in Fig.\ref{Fig-1DDecay_ECS}), and $c$ is the speed of light in cm/s. Note that the units of (cm$^{-1}$ atm$^{-1}$) indicate linear dependence of the decay rate on pressure, exactly as expected from a linear dependence of the collision rate on gas density. This linear relationship has been confirmed in our experiments conducted at various pressure values ranging from 0.5 to 1.5 atm.

The observed decoherence rate drops by more than an order of magnitude with the angular momentum of nitrogen molecules increasing from $J=8$ to $J=66$. At the lower end of the scale, $J<20$, which can be accessed at room temperature without the use of the centrifuge, our results are in good agreement with the known data from thermal ensembles\cite{Miller11}, shown in Fig.\ref{Fig-1DDecay_ECS} by black squares. At $J=10$, the decay rate corresponds to the exponential time constant of 62 ps, whereas it grows to 664 ps at $J=66$. Since the intensity of Raman signal scales as the square of the rotational coherence, this yields the coherence lifetime of 1.33 ns, or the equivalent of about 10 collisions. In contrast to ``slow rotors'' whose dynamics are altered by a single collision, super rotors are much more resilient to collisional relaxation.

We start the analysis of the observed behavior by first noting that our experimental technique cannot distinguish between the elastic and inelastic mechanisms of rotational decoherence. The latter mechanism, however, has been suggested as the main contributor to the decay of rotational coherence\cite{Strekalov00, Hartmann12}. It is therefore instructive to examine the utility of the energy corrected sudden approximation, which gives the recipe for calculating the rate of inelastic $J$-changing collisions, in describing our experimental findings. In ECS, the rate of transition from $J$ to $J'$ is given by the following expression:
\begin{eqnarray}\label{Eq_ECS}
    \gamma^{\textsc{esc}}_{J,J'} = (2J'+1) \; \exp \left( \frac{E_{J}-E_{J_>}}{k_{B}T} \right) \times \nonumber \\ \sum_{L} \left( \begin{array}{ccc} J & J' & L \\ 0 & 0 & 0 \end{array} \right)^{2} (2L+1) \frac{\Omega_{l_c,v_c} (J)}{\Omega_{l_c,v_c} (L)} \gamma_{L0},
\end{eqnarray}
where $E_{J}$ is the rotational energy (with $J_>$ denoting the largest value between $J$ and $J'$), (:::) is the Wigner $3J$ symbol, and $\Omega _{l_c,v_c}(J)$ is the correction factor introduced earlier. Given that the basic rate $\gamma _{L0}$ is known to fall off rather steeply with $L$ \cite{Brunner81}, we simplify Eq.\ref{Eq_ECS} by leaving only the single dominant term with $L=2$ in the sum. Because of the exponential gap factor in the first line, we also assume that the main contribution to the decay of rotational coherence comes from the downward transitions with $J'=J-2$. Noticing that the corresponding $3J$ symbol scales as $1/\sqrt{J}$ at large $J$, we arrive at the following simplified decay rate:
\begin{equation}\label{Eq_Gamma}
    \Gamma_{J\gg1} \approx \gamma^{\textsc{esc}}_{\substack{J,J-2 \\ (J\gg1)}} \approx \frac{15\gamma _{20}}{8\Omega_{l_c,v_c}(2)}\Omega_{l_c,v_c} (J)\equiv A\Omega_{l_c,v_c} (J),
\end{equation}
where all the factors which are independent on $J$ have been included in $A$. With only two fitting parameters, $A$ and $l_c$, this simple expression results in a reasonably good agreement with our experimental observations in the region of high $J$ values. This is demonstrated by fitting the data using Eq.\ref{Eq_Gamma} at $J>12$, i.e. above the visible bend in the curve predicted by the full ECS model and observed here similarly to a number of previous reports\cite{Martinsson93, Miller11, Kliewer12}. The fit (solid blue line in Fig.\ref{Fig-1DDecay_ECS}) corresponds to the characteristic interaction length $l_c=0.74\pm0.03$~\AA, in excellent agreement with the previously reported value of 0.75~\AA\cite{Martinsson93, Thumann97}.

To test the proposed simplified scaling law, we repeated the experiment at a higher temperature of $T=503$~K. The measured decay rates are shown with red circles in Fig.\ref{Fig-1DDecay_ECS}. Importantly, fitting the high-temperature data for $J>12$ with Eq.\ref{Eq_Gamma} yields the same characteristic length $l_c=0.74$~\AA, confirming the validity of the model. In accord with all previous observations of thermally accessible $J$ levels, increasing the temperature results in a slower decay rate, primarily due to the reduced gas density. Increasing the value of $J$ leads to the similar decrease of $\Gamma_{J}$, this time due to the growing adiabaticity of collisions. However, as can be seen from the inset in Fig.\ref{Fig-1DDecay_ECS}, the colder the ensemble the faster the growth of the adiabaticity factor $a$ with $J$. This fact explains the observed reversal of the temperature dependence between $J\approx 35$ and $J\approx 50$, where the molecules in a hot (and therefore more dilute) ensemble lose their rotational coherence faster than the denser cold ones. In this window of $J$'s, the value of $a$ is already above the adiabaticity threshold at $T=$294~K, yet still below it at $T=$503~K.

Black circles in the inset to Fig.\ref{Fig-1DDecay_ECS} mark the highest observed states of nitrogen at $T=300,500$ and 1100~K (also corresponding to the rotational levels containing 0.2\% of the total rotational population). One can see that, although the range of $J$ values can be extended by increasing the temperature, the maximum accessible adiabaticity parameter remains constant and rather low. Controlling molecular rotation with an optical centrifuge eliminates this limitation and enables one to cross over to and explore the adiabatic regime of rotational decoherence.

In summary, we have investigated the effect of ultrafast molecular rotation on the collision-induced rotational decoherence in the regime when the speed of molecular rotation exceeds the relative speed between the collision partners. We have found a satisfactory agreement with a simplified scaling law, in which the decay of rotational coherence depends on the degree of adiabaticity of the collision process. The demonstrated robustness of molecular super rotors against collisional relaxation opens interesting prospects of using these molecular objects in numerous schemes of coherent control based on rotational dynamics.

\begin{acknowledgements}
This work has been supported by the CFI, BCKDF and NSERC. We thank Ilya Sh. Averbukh for many stimulating discussions. We are also grateful to C.~J.~Kliewer for valuable comments on the effect of spatial diffusion and the existing data on nitrogen thermometry.
\end{acknowledgements}


\end{document}